\documentclass[numberedappendix]{emulateapj} 

\usepackage{natbib}
\usepackage{hyperref}
\hypersetup{colorlinks,citecolor=Blue,linkcolor=Red,urlcolor=Blue}
\usepackage{amsmath}
\usepackage{empheq}
\usepackage[usenames,dvipsnames]{color}

\allowdisplaybreaks 

\begin{document}
 
\title{Solar Obliquity Induced by Planet Nine}  
\author{Elizabeth Bailey, Konstantin Batygin, Michael E. Brown} 

\affil{Division of Geological and Planetary Sciences, California Institute of Technology, Pasadena, CA 91125} 
\email{ebailey@gps.caltech.edu}
 
\newcommand{\Ham}{\mathcal{H}}
\newcommand{\G}{\mathcal{G}}
\newcommand{\appropto}{\mathrel{\vcenter{\offinterlineskip\halign{\hfil$##$\cr\propto\cr\noalign{\kern2pt}\sim\cr\noalign{\kern-2pt}}}}}
\newcommand{\Poincare}{{Poincar$\acute{\rm{e}}$}}

\begin{abstract} 

The six-degree obliquity of the sun suggests that either an asymmetry was present in the solar system's formation environment, or an external torque has misaligned the angular momentum vectors of the sun and the planets. However, the exact origin of this obliquity remains an open question. \citet{BB16a} have recently shown that the physical alignment of distant Kuiper Belt orbits can be explained by a $5-20\,m_{\oplus}$ planet on a distant, eccentric, and inclined orbit, with an approximate perihelion distance of $\sim250\,$AU. Using an analytic model for secular interactions between Planet Nine and the remaining giant planets, here we show that a planet with similar parameters can naturally generate the observed obliquity as well as the specific pole position of the sun's spin axis, from a nearly aligned initial state. Thus, Planet Nine offers a testable explanation for the otherwise mysterious spin-orbit misalignment of the solar system.
\end{abstract} 

\maketitle

\section{Introduction} \label{sect1}

The axis of rotation of the sun is offset by six degrees from the invariable plane of the solar system \citep{SouSou12}. In contrast, planetary orbits have an RMS inclination slightly smaller than one degree\footnote{An exception to the observed orbital coplanarity of the planets is Mercury, whose inclination is subject to chaotic evolution \citep{Laskar94, BatyginHolmanMorby15}}, rendering the solar obliquity a considerable outlier. The origin of this misalignment between the sun's rotation axis and the angular momentum vector of the solar system has been recognized as a a longstanding question \citep{Kuip56, Trem91, heller93}, and remains elusive to this day.

With the advent of extensive exoplanetary observations, it has become apparent that significant spin-orbit misalignments are common, at least among transiting systems for which the stellar obliquity can be determined using the Rossiter-McLaughlin effect \citep{Rossiter, McLaugh}. Numerous such observations of planetary systems hosting hot Jupiters have revealed spin-orbit misalignments spanning tens of degrees \citep{Hebrard, Winn, Albrecht}, even including observations of retrograde planets \citep{Narita, Winn09, Bayliss, Winn11}. Thus, when viewed in the extrasolar context, the solar system seems hardly misaligned. However, within the framework of the nebular hypothesis, the expectation for the offset between the angular momentum vectors of the planets and sun is to be negligible, unless a specific physical mechanism induces a misalignment. Furthermore, the significance of the solar obliquity is supported by the contrasting relative coplanarity of the planets.

Because there is no directly observed stellar companion to the sun (or any other known gravitational influence capable of providing an external torque on the solar system sufficient to produce a six-degree misalignment over its multi-billion-year lifetime \citealt{heller93}), virtually all explanations for the solar obliquity thus far have invoked mechanisms inherent to the nebular stage of evolution. In particular, interactions between the magnetosphere of a young star and its protostellar disk can potentially lead to a wide range of stellar obliquities while leaving the coplanarity of the tilted disk intact \citep{Lai11}. Yet another possible mechanism by which the solar obliquity could be attained in the absence of external torque is an initial asymmetry in the mass distribution of the protostellar core. Accordingly, asymmetric infall of turbulent protosolar material has been proposed as a mechanism for the sun to have acquired an axial tilt upon formation \citep{Bate, Fielding15}. However, the capacity of these mechanisms to overcome the re-aligning effects of accretion, as well as gravitational and magnetic coupling, remains an open question \citep{Lai11, Spalding14, Spalding15}.

In principle, solar obliquity could have been excited through a temporary, extrinsic gravitational torque early in the solar system's lifetime. That is, an encounter with a passing star or molecular cloud could have tilted the disc or planets with respect to the sun \citep{heller93,Adams10}. Alternatively, the sun may have had a primordial stellar companion, capable of early star-disc misalignment \citep{Bat12,Spalding14,Lai14}. To this end, ALMA observations of misaligned disks in stellar binaries \citep{Jensen, Williams} have provided evidence for the feasibility of this effect.  Although individually sensible, a general qualitative drawback of all of the above mechanisms is that they are only testable when applied to the extrasolar population of planets, and it is difficult to discern which (if any) of the aforementioned processes operated in our solar system. 

Recently, \citet{BB16a} determined that the spatial clustering of the orbits of Kuiper Belt objects with semi-major axis $a\gtrsim250\,$AU can be understood if the solar system hosts an additional $m_{9}=5-20\,m_{\oplus}$ planet on a distant, eccentric orbit. Here, we refer to this object as Planet Nine. The orbital parameters of this planet reside somewhere along a swath of parameter space spanning hundreds of AU in semi-major axis, significant eccentricity, and tens of degrees of inclination, with a perihelion distance of roughly $q_{9}\sim250\,$AU \citep{BB16b}. In this work, we explore the possibility that this distant, planetary-mass body is fully or partially responsible for the peculiar spin axis of the sun. 

Induction of solar obliquity of some magnitude is an inescapable consequence of the existence of Planet Nine. That is, the effect of a distant perturber residing on an inclined orbit is to exert a mean-field torque on the remaining planets of the solar system, over a timespan of $\sim4.5$ Gyr. In this manner, the gravitational influence of Planet Nine induces precession of the angular momentum vectors of the sun and planets about the total angular momentum vector of the solar system. Provided that angular momentum exchange between the solar spin axis and the planetary orbits occurs on a much longer timescale, this process leads to a differential misalignment of the sun and planets. Below, we quantify this mechanism with an eye towards explaining the tilt of the solar spin axis with respect to the orbital angular momentum vector of the planets.

The paper is organized as follows. Section (\ref{sect2}) describes the dynamical model. We report our findings in section (\ref{sect3}). We conclude and discuss our results in section (\ref{sect4}). Throughout the manuscript, we adopt the following notation. Similarly named quantities (e.g. $a$, $e$, $i$) related to Planet Nine are denoted with a subscript ``9", whereas those corresponding to the Sun's angular momentum vector in the inertial frame are denoted with a tilde. Solar quantities measured with respect to the solar system's invariable plane are given the subscript $\odot$.

\begin{figure}
\includegraphics[width=0.48\textwidth]{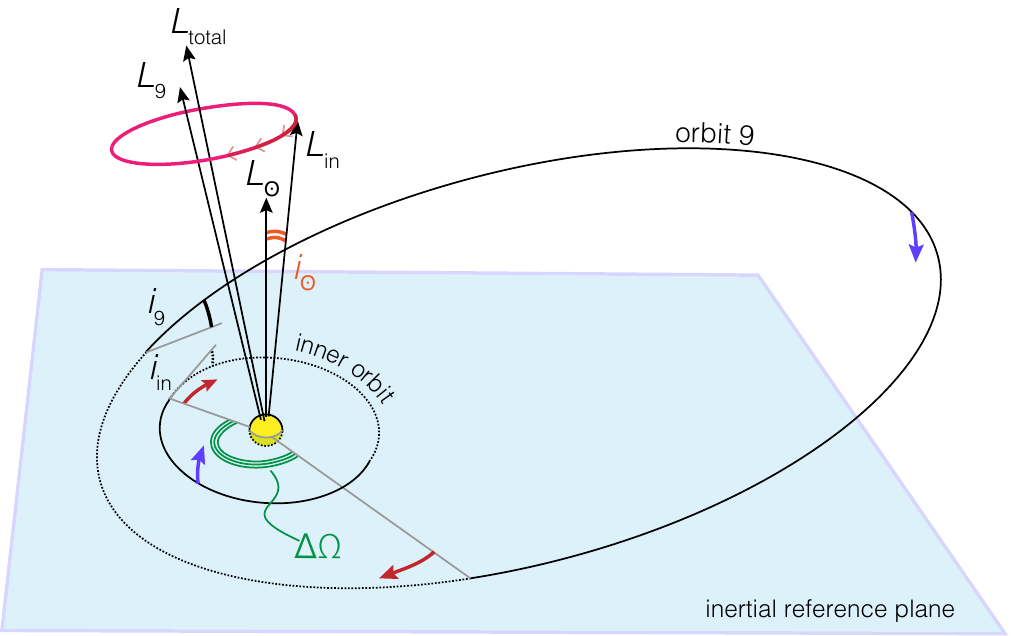}
\caption{Geometric setup of the dynamical model. The orbits of the planets are treated as gravitationally interacting rings. All planets except Planet Nine are assumed to have circular, mutually coplanar orbits, and are represented as a single inner massive wire. The sun is shown as a yellow sphere, and elements are not to scale. Black, grey, and dotted lines are respectively above, on, and below the inertial reference plane. The pink arrows demonstrate the precession direction of the angular momentum vector of the inner orbit, $L_{\text{in}}$, around the total angular momentum vector of the solar system $L_{\text{total}}$. Red and blue arrows represent the differential change in longitudes of ascending node of the orbits and inclination, respectively. Although not shown in the figure, the tilting of the oblate sun is modeled as the tilting of an inner test ring. Over the course of 4.5 billion years, differential precession of the orbits induces a several-degree solar obliquity with respect to the final plane of the planets.} 
\label{setup}
\end{figure}

\section{Dynamical Model} \label{sect2}
To model the long-term angular momentum exchange between the known giant planets and Planet Nine, we employ secular perturbation theory. Within the framework of this approach, Keplerian motion is averaged out, yielding semi-major axes that are frozen in time. Correspondingly, the standard $N-$planet problem is replaced with a picture in which $N$ massive wires (whose line densities are inversely proportional to the instantaneous orbital velocities) interact gravitationally \citep{MD99}. Provided that no low-order commensurabilities exist among the planets, this method is well known to reproduce the correct dynamical evolution on timescales that greatly exceed the orbital period \citep{Mardling07, Li14}.

In choosing which flavor of secular theory to use, we must identify small parameters inherent to the problem. Constraints based upon the critical semi-major axis beyond which orbital alignment ensues in the distant Kuiper belt, suggest that Planet Nine has an approximate perihelion distance of $q_{9}\sim250\,$AU and an appreciable eccentricity $e_{9}\gtrsim0.3$ \citep{BB16a, BB16b}. Therefore, the semi-major axis ratio $(a/a_{9})$ can safely be assumed to be small. Additionally, because solar obliquity itself is small and the orbits of the giant planets are nearly circular, here we take $e=0$ and $\sin(i)\ll1$. Under these approximations, we can expand the averaged planet-planet gravitational potential in small powers of $(a/a_{9})$, and only retain terms of leading order in $\sin(i)$.

In principle, we could self-consistently compute the interactions among all of the planets, including Planet Nine. However, because the fundamental secular frequencies that characterize angular momentum exchange among the known giant planets are much higher than that associated with Planet Nine, the adiabatic principle \citep{Henrard1982,Neishtadt1984} ensures that Jupiter, Saturn, Uranus and Neptune will remain co-planar with each-other throughout the evolutionary sequence (see e.g. \citealt{Bat11,Bat12} for a related discussion on perturbed self-gravitating disks). As a result, rather than modeling four massive rings individually, we may collectively replace them with a single circular wire having semi-major axis $a$ and mass $m$, and possessing equivalent total angular momentum and moment of inertia:
\begin{align}
m\,\sqrt{a} &= \sum_{j} m_j\,\sqrt{a_j} \nonumber \\
m\,a^2 &= \sum_{j} m_j\,a_j^2,
\label{singlewire}
\end{align}
where the index $j$ runs over all planets. The geometric setup of the problem is shown in Figure (\ref{setup}).

\begin{figure*}
\includegraphics[width=1\textwidth]{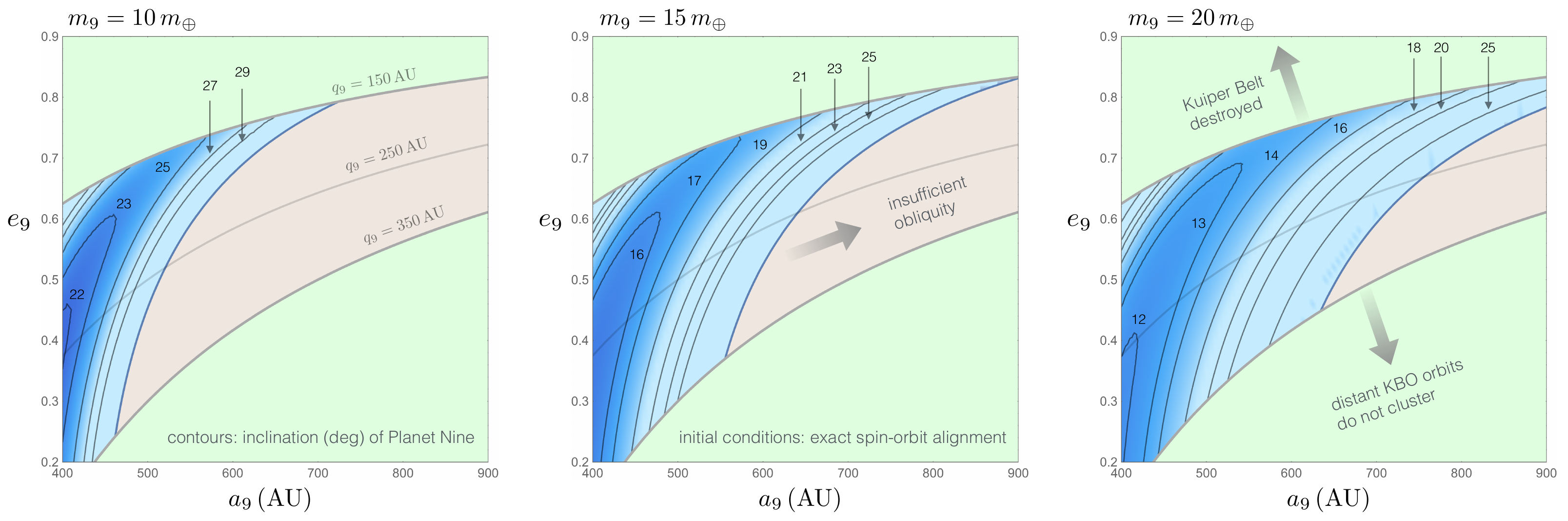}
\caption{Parameters of Planet Nine required to excite a spin-orbit misalignment of $i_{\odot}=6\,\deg$ over the lifetime of the solar system, from an initially aligned state. Contours in $a_9$-$e_9$ space denote $i_{9}$, required to match the present-day solar obliquity. Contour labels are quoted in degrees. The left, middle, and right panels correspond to $m_9=10,$ $15$, and $20\,m_{\oplus}$ respectively. Due to independent constraints stemming from the dynamical state of the distant Kuiper belt, only orbits that fall in the $150<q_9<350\,$AU range are considered. The portion of parameter space where a solar obliquity of $i_{\odot}=6\,\deg$ cannot be attained are obscured with a light-brown shade.} 
\label{inclination}
\end{figure*}

To quadrupole order, the secular Hamiltonian governing the evolution of two interacting wires is \citep{Kaula62, Mardling10}:
\begin{align}
\mathcal{H} &=\frac{\G}{4}\frac{m\, m_{9}}{a_{9}} \bigg( \frac{a}{a_{9}} \bigg)^2 \frac{1}{\varepsilon_9^3} \bigg[ \frac{1}{4} \big(3 \cos^2(i) -1 \big) \big(3 \cos^2(i_{9}) -1 \big) \nonumber \\
&+\frac{3}{4} \sin(2\,i)\,\sin(2\,i_{9})\,\cos(\Omega - \Omega_{9}) \bigg],
\label{HammyKep}
\end{align}
where $\Omega$ is the longitude of ascending node and $\varepsilon_9 = \sqrt{1-e_{9}^2}$. Note that while the eccentricities and inclinations of the known giant planets are assumed to be small, no limit is placed on the orbital parameters of Planet Nine. Moreover, at this level of expansion, the planetary eccentricities remain unmodulated, consistent with the numerical simulations of \cite{BB16a,BB16b}, where the giant planets and Planet Nine are observed to behave in a decoupled manner.

Although readily interpretable, Keplerian orbital elements do not constitute a canonically conjugated set of coordinates. Therefore, to proceed, we introduce \Poincare\ action-angle coordinates:
\begin{align}
\Gamma &= m \sqrt{\G\,M_{\odot}\,a} \nonumber \\
\Gamma_{9} &= m_{9} \sqrt{\G\,M_{\odot}\,a_{9}} \, \varepsilon_9 \nonumber \\
Z &= \Gamma \big(1- \cos(i) \big)    &&z=-\Omega \nonumber \\
Z_{9} &= \Gamma_{9} \big(1- \cos(i_{9}) \big)  &&z=-\Omega_{9}.
\label{AAcoords}
\end{align}
Generally, the action $Z$ represents the deficit of angular momentum along the $\hat{\bf{k}}-$axis, and to leading order, $i \approx \sqrt{2 Z/\Gamma}$. Accordingly, dropping higher-order corrections in $i$, expression (\ref{HammyKep}) takes the form:
\begin{align}
\mathcal{H} &=\frac{\G}{4}\frac{m\, m_{9}}{a_{9}} \bigg( \frac{a}{a_{9}} \bigg)^2 \frac{1}{\epsilon_{9}^3} \bigg[ \frac{1}{4}\bigg(2-\frac{6 Z}{\Gamma} \bigg) \bigg(3\bigg( 1-\frac{Z_{9}}{\Gamma_{9}} \bigg)^2-1 \bigg) \nonumber \\ 
&+ 3 \bigg(1 - \frac{Z_{9}}{\Gamma_{9}} \bigg) \sqrt{1-\frac{Z_{9}}{2\Gamma_{9}}} \sqrt{\frac{2 Z}{\Gamma} \frac{2 Z_{9}}{\Gamma_{9}}} \cos(z-z_{9}) \bigg].
\label{HammyAA}
\end{align}
Application of Hamilton's equations to this expression yields the equations of motion governing the evolution of the two-ring system. However, we note that action-angle variables (\ref{AAcoords}) are singular at the origin, so an additional, trivial change to Cartesian counterparts of \Poincare\ coordinates is required to formulate a practically useful set of equations \citep{Morbybook}. This transformation is shown explicitly in the Appendix. 

\begin{figure*}
\includegraphics[width=1\textwidth]{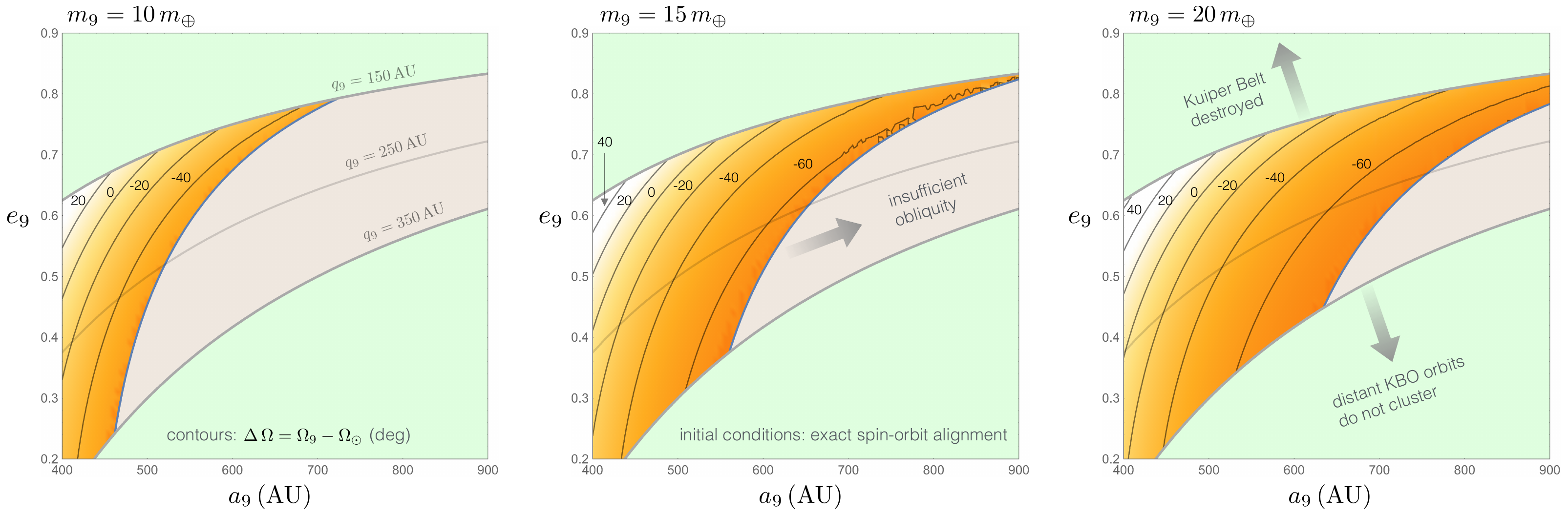}
\caption{This set of plots depict the same parameter space as in Figure (\ref{inclination}), but the contours represent the longitude of ascending node of Planet Nine, relative to that of the Sun, $\Delta\,\Omega$. As before the values are quoted in degrees.} 
\label{node}
\end{figure*}

To complete the specification of the problem, we also consider the torque exerted on the sun's spin axis by a tilting solar system. Because the sun's angular momentum budget is negligible compared to that of the planets, its back-reaction on the orbits can be safely ignored. Then, the dynamical evolution of its angular momentum vector can be treated within the same framework of secular theory, by considering the response of a test ring with semi-major axis \citep{Spalding14, Spalding15}:
\begin{align}
\tilde{a}=\left[\frac{16\,\omega^2\,k_2^2\,R^6}{9\,I^2\,\G\,M_{\odot}}\right]^{1/3},
\label{astar}
\end{align}
where $\omega$ is the rotation frequency, $k_2$ is the Love number, $R$ is the solar radius, and $I$ is the moment of inertia.

Because we are primarily concerned with main-sequence evolution, here we adopt $R=R_{\odot}$ and model the interior structure of the sun as a $n=3$ polytrope, appropriate for a fully radiative body \citep{Chandrasek39}. Corresponding values of moment of inertia and Love number are $I=0.08$ and $k_2=0.01$ respectively 
\citep{BatAd13}. The initial rotation frequency is assumed to correspond to a period of $2\pi/\omega=10\,$days and is taken to decrease as $\omega\propto1/\sqrt{t}$, in accord with the Skumanich relation \citep{GalBou13}. 

Defining scaled actions $\tilde{\Gamma}= \sqrt{\G\,M_{\odot}\,\tilde{a}}$ and $\tilde{Z}= \tilde{\Gamma}(1-\cos(\tilde{i}))$ and scaling the Hamiltonian itself in the same way, we can write down a Hamiltonian that is essentially analogous to equation (\ref{HammyAA}), which governs the long-term spin axis evolution of the Sun:
\begin{align}
\tilde{\mathcal{H}} &=\sum_{j}\left(\frac{\G\,m_j}{4\,a_j^3}\right)\tilde{a}^2 \bigg[\frac{3 \tilde{Z}}{\tilde{\Gamma}}  + \frac{3}{4} \sqrt{\frac{2 \tilde{Z}}{\tilde{\Gamma}} \frac{2 Z}{\Gamma}} \cos(\tilde{z}-z) \bigg].
\label{HammySun}
\end{align}
Note that contrary to equation (\ref{HammyAA}), here we have assumed small inclinations for both the solar spin axis and the planetary orbits. This assumption transforms the Hamiltonian into a form equivalent to the Lagrange-Laplace theory, where the interaction coefficients have been expanded as hypergeometric series, to leading order in semi-major axis ratio \citep{MD99}. Although not particularly significant in magnitude, we follow the evolution of the solar spin axis for completeness. 

Quantitatively speaking, there are two primary sources of uncertainty in our model. The first is the integration timescale. Although the origin of Planet Nine is not well understood, its early evolution was likely affected by the presence of the solar system's birth cluster \citep{Izidoro2015, LiAd16}, meaning that Planet Nine probably attained its final orbit within the first $\sim100\,$Myr of the solar system's lifetime. Although we recognize the $\sim2\%$ error associated with this ambiguity, we adopt an integration timescale of $4.5\,$Gyr for definitiveness. 

A second source of error stems from the fact that the solar system's orbital architecture almost certainly underwent a instability-driven transformation sometime early in its history \citep{Tsiganis,NesvornyMorby}. Although the timing of the onset of instability remains an open question \citep{Levison11, KaibChambers}, we recognize that failure of our model to reflect this change in $a$ and $m$ (through equation \ref{singlewire}) introduces a small degree of inaccuracy into our calculations. Nevertheless, it is unlikely that these detailed complications constitute a significant drawback to our results.

\section{Results} \label{sect3}

The Sun's present-day inclination with respect to the solar system's invariable plane\footnote{Although we refer to the instantaneous plane occupied by the wire with parameters $a$ and $m$ as the invariable plane, in our calculations, this plane is not actually invariable. Instead, it slowly precesses in the inertial frame.} \citep{SouSou12} is almost exactly $i_{\odot}=6\,\deg$. Using this number as a constraint, we have calculated the possible combinations of $a_{9}$, $e_{9}$ and $i_{9}$ for a given $m_{9}$, that yield the correct spin-orbit misalignment after $4.5\,$Gyr of evolution. For this set of calculations, we adopted an initial condition in which the sun's spin axis and the solar system's total angular momentum vector were aligned. 

The results are shown in Figure (\ref{inclination}). For three choices of $m_9=10$, $15$, and $20\,m_{\oplus}$, the Figure depicts contours of the required $i_{9}$ in $a_{9}-e_{9}$ space. Because Planet Nine's perihelion distance is approximately $q_{9}\sim250\,$AU, we have only considered orbital configurations with $150<q_{9}<350\,$AU. Moreover, within the considered locus of solutions, we neglect the region of parameter space where the required solar obliquity cannot be achieved within the lifetime of the solar system. This section of the graph is shown with a light brown shade in Figure (\ref{inclination}).

For the considered range of $m_9$, $a_9$ and $e_9$, characteristic inclinations of $i_9\sim15-30\,\deg$ are required to produce the observed spin-orbit misalignment. This compares favorably with the results of \cite{BB16b}, where a similar inclination range for Planet Nine is obtained from entirely different grounds. However, we note that the constraints on $a_9$ and $e_9$ seen in Figure (\ref{inclination}) are somewhat more restrictive than those in previous works. In particular, the illustrative $m_9=10\,m_{\oplus}$, $a_9=700\,$AU, $e_9=0.6$ perturber considered by \citet{BB16a}, as well as virtually all of the ``high-probability" orbits computed by \cite{BB16b} fall short of exciting 6 degrees of obliquity from a strictly coplanar initial configuration. Instead, slightly smaller spin-orbit misalignments of $i_{\odot}\sim3-5\,\deg$ are typically obtained. At the same time, we note that the lower bound on the semi-major axis of Planet Nine quoted in \cite{BB16b} is based primarily on the comparatively low perihelia of the unaligned objects, rather than the alignment of distant Kuiper belt objects, constituting a weaker constraint. 

\begin{figure*}
\includegraphics[width=1\textwidth]{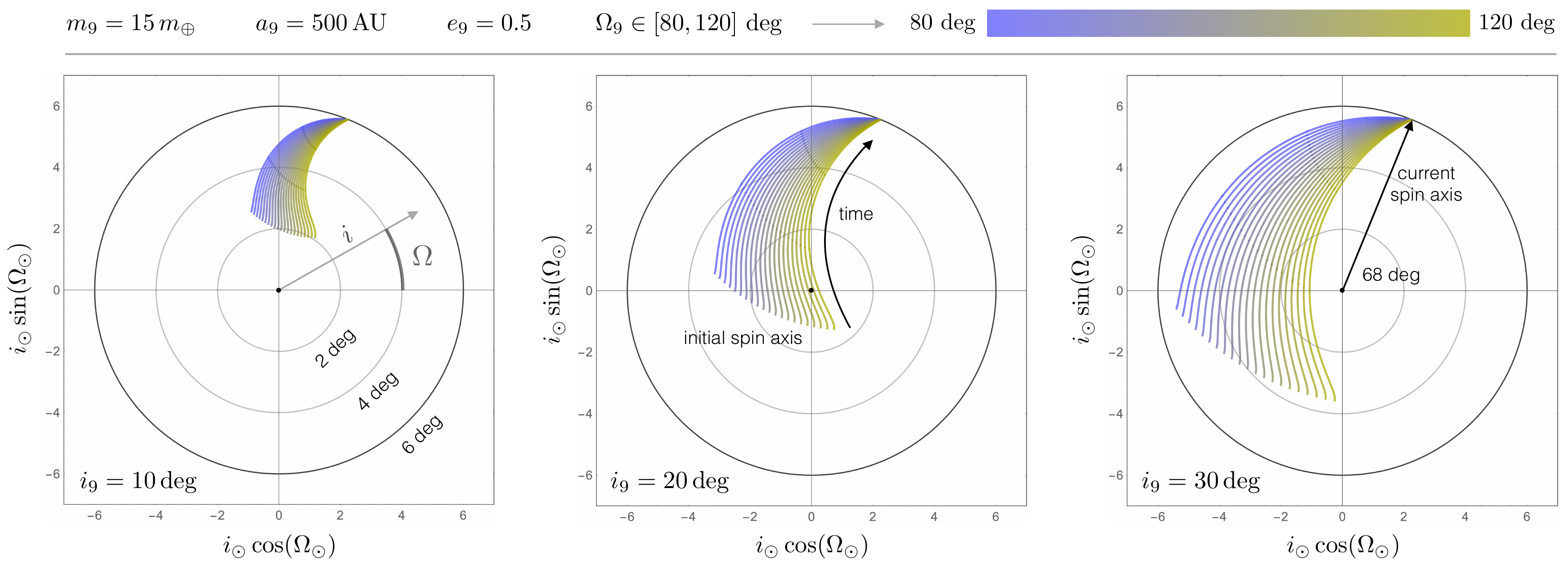}
\caption{Illustrative evolution tracks of the solar spin axis, measured with respect to the instantaneous invariable plane. The graphs are shown in polar coordinates, where $i_{\odot}$ and $\Omega_{\odot}$ represent the radial and angular variables respectively. The integrations are initialized with the Sun's present-day configuration ($i_{\odot}=6\,\deg$, $\Omega_{\odot}=68\,\deg$), and are performed backwards in time. For Planet Nine, parameters of $m_9=15\,m_{\oplus}$, $a_9=500\,$AU, $e_9=0.5$ are adopted throughout. Meanwhile, the left, middle, and right panels show results corresponding to $i_9=10,$ $20,$ and $30\,\deg$ respectively. The present-day and longitude of ascending node of Planet Nine is assumed to lie in the range $80<\Omega_9<120\,\deg$ and is represented by the color of the individual evolution tracks.} 
\label{backwards}
\end{figure*}

An equally important quantity as the solar obliquity itself, is the solar longitude of ascending node\footnote{The quoted value is measured with respect to the invariable plane, rather than the ecliptic.} $\Omega_{\odot}\simeq68\,\deg$. This quantity represents the azimuthal orientation of the spin axis and informs the direction of angular momentum transfer within the system. While the angle itself is measured from an arbitrary reference point, the difference in longitudes of ascending node $\Delta\,\Omega=\Omega_{9}-\Omega_{\odot}$ is physically meaningful, and warrants examination. 

Figure (\ref{node}) shows contours of $\Delta\,\Omega$ within the same parameter space as Figure (\ref{inclination}). Evidently, the representative range of the relative longitude of ascending node is $\Delta\,\Omega\sim-60-40\,\deg$, with the positive values coinciding with high eccentricities and low semi-major axes. Therefore, observational discovery of Planet Nine with a correspondent combination of parameters $a_9$, $e_9$, $i_9$, and $\Delta\Omega$ depicted anywhere on an analog of Figures (\ref{inclination}) and (\ref{node}) constructed for the specific value of $m_9$, would constitute formidable evidence that Planet Nine is solely responsible for the peculiar spin axis of the sun. On the contrary, a mismatch of these parameters relative to the expected values, would imply that Planet Nine has merely modified the sun's spin axis by a significant amount.

Although $\Omega_{9}$ is not known, Planet Nine's orbit is theoretically inferred to reside in approximately the same plane as the distant Kuiper belt objects, whose longitudes of ascending node cluster around $\langle \Omega \rangle=113\pm13\,\deg$ \citep{BB16a}. Therefore, it is likely that $\Omega_{9}\simeq\langle \Omega \rangle$, implying that $\Delta\,\Omega\simeq45\,\deg$. Furthermore, the simulation suite of \cite{BB16b} approximately constrains Planet Nine's longitude of ascending node to the range $\Omega_{9} \simeq 80 - 120\,\deg$, yielding $12<\Delta\,\Omega<52\,\deg$ as an expected range of solar spin axis orientations.  

If we impose the aforementioned range of $\Delta\,\Omega$ as a constraint on our calculations, Figure (\ref{node}) suggests that $a_9\lesssim500\,$AU and $e_9\gtrsim0.4$. Although not strictly ruled out, orbits that fall in this range are likely to be incompatible with the observed orbital architecture of the distant Kuiper belt. As a result, we speculate that either (I) Planet Nine does not reside in the same plane as the distant Kuiper belt objects it shepherds, or (II) our adopted initial condition where the sun's primordial angular momentum vector coincides exactly with that of the solar system is too restrictive. Of these two possibilities, the latter is somewhat more likely.

While a null primordial obliquity is a sensible starting assumption, various theoretical studies have demonstrated that that substantial spin-orbit misalignments can be excited in young planetary systems \citep{Lai11, Bat12, Lai14, Spalding14, Spalding15, Fielding15}, with substantial support coming from existing exoplanet data \citep{Huber13, WinnFab}. At the same time, the recent study of \cite{Spalding16} has suggested that a fraction of multi-transiting exoplanet systems would be rendered unstable if their host stars had obliquities as large as that of the Sun, and instead inclinations as small as $1-2\,\deg$ are more typical. Accordingly, it is sensible to suppose that the initial obliquity of the sun was not too different from the RMS inclination of the planets $i_{\rm{RMS}}\sim1\,\deg$. 

To examine this possibility, we considered whether a Planet Nine with $q_{9}=250\,$AU and $\Delta\,\Omega$ within the quoted range is consistent with a primordial solar obliquity of order $\sim1-2\,\deg$. As an illustrative example, we adopted $a_{9}=500\,$AU, $e_{9}=0.5$, $m_{9}=15\,m_{\odot}$, and evolved the system backwards in time. Because Hamiltonian (\ref{HammyAA}) is integrable, a present-day combination of parameters maps onto a unique primordial state vector. 

The calculations were performed for $i_{9}=10$, $20$, and $30\,\deg$, and the results are shown in Figure (\ref{backwards}). Specifically, the panels depict a polar representation of the sun's spin axis evolution tracks measured from the instantaneous invariable plane, such that the origin represents an exactly aligned configuration. The color of each curve corresponds to a current value of $\Omega_{9}$. Evidently, for the employed set of parameters, the calculations yield a primordial inclination range of $i_{\odot}\simeq1-6\,\deg$. Intriguingly, the specific choice of $i_{9}=20\deg$, and $\Omega_{9}\simeq\langle \Omega \rangle$ yields the lowest spin-orbit misalignment, that is consistent with $i_{\rm{RMS}}$. Therefore, we conclude that the notion of Planet Nine as a dominant driver of solar obliquity is plausible.

\section{Discussion} \label{sect4}

Applying the well-established analytic methods of secular theory, we have demonstrated that a solar obliquity of order several degrees is an expected observable effect of Planet Nine. Moreover, for a range of masses and orbits of Planet Nine that are broadly consistent with those predicted by \cite{BB16a, BB16b}, Planet Nine is capable of reproducing the observed solar obliquity of $6$ degrees, from a nearly coplanar configuration. The existence of Planet Nine therefore provides a tangible explanation for the spin-orbit misalignment of the solar system.

Within the context of the Planet Nine hypothesis, a strictly null tilt of the solar spin-axis is disallowed. However, as already mentioned above, in addition to the long-term gravitational torques exerted by Planet Nine, numerous other physical processes are thought to generate stellar obliquities (see e.g. \citealt{CridaBat} and the references therein). A related question then, concerns the role of Planet Nine with respect to every other plausible misalignment mechanism. Within the context of our model, this question is informed by the present-day offset between the longitudes of ascending node of Planet Nine and the Sun, $\Delta\,\Omega$. Particularly, if we assume that the solar system formed in a configuration that was strictly co-planar with the sun's equator, the observable combination of the parameters $m_9,a_9,e_9,i_9$ maps onto a unique value of the observable parameter $\Delta\,\Omega$. 

Importantly, our calculations suggest that if the orbit of Planet Nine resides in approximately the same plane as the orbits of the $a\gtrsim250\,$AU Kuiper belt objects (which inform the existence of Planet Nine in the first place), then the inferred range of $\Delta\,\Omega$ and Planet Nine's expected orbital elements are incompatible with an exactly co-linear initial state of the solar spin axis. Instead, backwards integrations of the equations of motion suggest that a primordial spin-orbit misalignment of the same order as the RMS spread of the planetary inclination ($i\sim1\,\deg$) is consistent with the likely orbital configuration of Planet Nine. In either case, our results contextualize the primordial solar obliquity within the emerging extrasolar trend of small spin-orbit misalignments in flat planetary systems \citep{MorWin}, and bring the computed value closer to the expectations of the nebular hypothesis. However, we note that at present, the range of unconstrained parameters also allows for evolutionary sequences in which Planet Nine's contribution does not play a dominant role in exciting the solar obliquity.

The integrable nature of the calculations performed in this work imply that observational characterization of Planet Nine's orbit will not only verify the expansion of the solar system's planetary album, but will yield remarkable new insights into the state of the solar system, at the time of its formation. That is, if Planet Nine is discovered in a configuration that contradicts a strictly aligned initial condition of the solar spin axis and planetary angular momentum, calculations of the type performed herein can be used to deduce the true primordial obliquity of the sun. In turn, this information can potentially constrain the mode of magnetospheric interactions between the young sun and the solar nebula \citep{Konigl91, Lai11, Spalding15}, as well as place meaningful limits on the existence of a putative primordial stellar companion of the sun \citep{Bat12,XiangGruessPapaloizou}.

Finally, this work provides not only a crude test of the likely parameters of Planet Nine, but also a test of the viability of the Planet Nine hypothesis. By definition, Planet Nine is hypothesized to be a planet having parameters sufficient to induce the observed orbital clustering of Kuiper belt objects with semi-major axis $a>250$ AU \citep{BB16a}. According to this definition, Planet Nine must occupy a narrow swath in $a-e$ space such that $q_{9}\sim250\,$AU, and its mass must reside in the approximate range $m_9=5-20\,m_{\oplus}$. If Planet Nine were found to induce a solar obliquity significantly higher than the observed value, the Planet Nine hypothesis could be readily rejected. Instead, here we have demonstrated that, over the lifetime of the solar system, Planet Nine typically excites a solar obliquity that is similar to what is observed, giving additional credence to the Planet Nine hypothesis.

\acknowledgments
\textbf{Acknowledgments}. We are grateful to Chris Spalding and Roger Fu for useful discussions. During the review of this paper, we have become aware that \cite{Gomes16} have reached similar conclusions simultaneously and independently.

\begin{appendix}

To octupole order in $(a/a_9)$, the full Hamiltonian governing the secular evolution of a hierarchical triple is \citep{Kaula62,Mardling10}:
\begin{align*}
\mathcal{H}&=-\frac{1}{4} \frac{\G\,\mu\,m_9}{a_9} \bigg( \frac{a}{a_9}\bigg)^2 \frac{1}{\varepsilon_9^3} \Bigg[ \bigg(1+\frac{3}{2}e^2\bigg)\frac{1}{4}\bigg(3 \cos(i) -1 \bigg)\bigg(3 \cos(i_9) -1\bigg) \\ &+ \frac{15}{14} e^2 \sin^2 (i) \cos(2\omega) + \frac{3}{4} \sin (2i) \sin (2 i_9) \cos(\Omega - \Omega_9) + \frac{3}{4} \sin^2(i) \sin^2 (i_9) \cos(2\Omega-2\Omega_9 )  \Bigg],
\end{align*}
where elements without a subscript refer to the inner body, and elements with subscript $9$ refer to the outer body, in this case Planet Nine. Here $\mu = (M_{\odot} m)/(M_{\odot}+m) \approx m$, and $\varepsilon_9$ is equal to $\sqrt{1-e_9^2}$. 

To attain integrability, we drop the Kozai harmonic because comparatively rapid perihelion precession of the known giant planets' orbits ensures that libration of $\omega$ is not possible \citep{BatyginTsiganisMorby2011}. Because the eccentricities of the known giant planets are small, we adopt $e=0$ for the inner orbit. Additionally, because the inclination of the inner orbit is presumed to be small throughout the evolutionary sequence, we neglect the higher-order $\cos(2\Omega-2\Omega_{9})$ harmonic, because it is proportional to $\sin^2(i)\ll\sin(2i)\ll1$. 

Keeping in mind the trigonometric relationship $\sin i = \sqrt{1-\cos^2 i}$, and adopting canonical \Poincare\ action-angle variables given by equation (\ref{AAcoords}), the Hamiltonian takes the approximate form
\begin{align*}
\mathcal{H}&=-\frac{1}{4} \frac{\G\,m\,m_9}{a_9} \bigg( \frac{a}{a_9}\bigg)^2 \frac{1}{\varepsilon^3} \Bigg[\frac{1}{4} \Bigg( 3 \bigg( 1-\frac{Z}{\Gamma} \bigg) ^2 -1 \Bigg) \Bigg( 3 \bigg(1-\frac{Z_9}{\Gamma_9}\bigg)^2 -1 \Bigg)  \\ 
&+ \frac{3}{4} \Bigg(2\bigg(1-\frac{Z}{\Gamma}\bigg)\sqrt{1-\bigg( 1- \frac{Z}{\Gamma} \bigg)^2} \Bigg) \Bigg(2\bigg(1-\frac{Z_9}{\Gamma_9}\bigg)\sqrt{1-\bigg( 1- \frac{Z_9}{\Gamma_9}
	 \bigg)^2} \Bigg) \cos{(z-z_9)}  \Bigg].
\end{align*}
Because the inner orbit has small inclination, it is suitable to expand $\mathcal{H}$ to leading order in $Z$. This yields the Hamiltonian given in equation (\ref{HammyAA}).
	
Since Hamiltonian (\ref{HammyAA}) possesses only a single degree of freedom, the Arnold-Liouville theorem \citep{Arnold1963} ensures that by application of the Hamilton-Jacobi equation, $\mathcal{H}$ can be cast into a form that only depends on the actions. Then, the entirety of the system's dynamics is encapsulated in the linear advance of cyclic angles along contours defined by the constants of motion \citep{Morbybook}. Here, rather than carrying out this extra step, we take the more practically simple approach of numerically integrating the equations of motion, while keeping in mind that the resulting evolution is strictly regular. 

The numerical evaluation of the system's evolution can be robustly carried out after transforming the Hamiltonian to nonsingular Poincar\'{e} Cartesian coordinates
\begin{align*}
x&=\sqrt{2Z}\cos{(z)} & y &=\sqrt{2Z}\sin{(z)}\\
x_9&= \sqrt{2Z_9}\cos{(z_9)} & y_9&=\sqrt{2Z_9}\sin{(z_9)}.
\end{align*}  
Then, the truncated and expanded Hamiltonian (\ref{HammyAA}) becomes
\begin{align*}
\mathcal{H}&=-\frac{1}{4} \frac{\G\,m\, m_9}{a_9} \bigg( \frac{a}{a_9}\bigg)^2 \frac{1}{\varepsilon_9^3} \Bigg[  \frac{1}{4} \bigg( 2-\frac{6 }{\Gamma}\bigg(\frac{x^2 + y^2}{2} \bigg) \bigg) \bigg( 3 \bigg( 1-\frac{1}{\Gamma_9} \bigg( \frac{x_9^2 + y_9^2}{2} \bigg) \bigg)^2 -1 \bigg)  \\
&+ 3 \bigg( 1-\frac{1}{\Gamma_9} \bigg( \frac{x_9^2 + y_9^2}{2} \bigg) \bigg) \sqrt{1-\frac{1}{2 \Gamma_9} \bigg( \frac{x_9^2 + y_9^2}{2} \bigg) } \sqrt{\frac{1}{\Gamma \Gamma_9}}\Big( xx_9 + yy_9 \Big)  \Bigg].
\end{align*}
Explicitly, Hamilton's equations $dx/dt=-\partial \mathcal{H} / \partial y$, $dy/dt=\partial \mathcal{H} / \partial x$ take the form:
\begin{align*}
\frac{dx}{dt}& = \frac{a^2 \G \,m\,m_9}{4\,a_9^3\,\epsilon_9^3} \Bigg( \frac{3y_9 \big(2\Gamma_9 - x_9^2 - y_9^2 \big)}{4\Gamma_9} \sqrt{\frac{4\Gamma_9 - x_9^2 - y_9^2}{\Gamma \Gamma_9^2}} + \frac{3y}{2\Gamma} \bigg( 1- \frac{3 \big( 2 \Gamma_9 - x_9^2 - y_9^2 \big)^2}{4 \Gamma_9^2} \bigg)  \Bigg)\\
\frac{\partial y }{\partial t}&= \frac{3\,a^2\,\G\, m\,m_9}{32\,a_9^3\,\Gamma\,\Gamma_9^2\,\epsilon_9^3} \Bigg(  2 x_9 \sqrt{\Gamma\big( 4\Gamma_9 -x_9^2 - y_9^2\big)} \big( x_9^2 + y_9^2 -2\Gamma_9 \big) + x \Big( 8\Gamma_9^2 + 3x_9^4 - 12\Gamma_9y_9^2 + 3 y_9^4 + 6 x_9^2 \Big( y_9^2 -2\Gamma_9 \Big) \Big) \Bigg)\\
\frac{\partial x_9}{\partial t} &= \frac{3\,a^2\,\G\,m\,m_9}{16\,a_9^3\,\Gamma_9^2\,\epsilon_9^3} \Bigg( -2y_9\Big( xx_9 + y y_9 \Big)\sqrt{\frac{4\Gamma_9 - x_9^2 -y_9^2}{\Gamma}} + y \Big( 2\Gamma_9-x_9^2 -y_9^2  \Big) \sqrt{\frac{4\Gamma_9 - x_9^2 -y_9^2}{\Gamma}}  \\
&+\frac{1}{\Gamma} y_9 \Big( 2\Gamma - 3x^2 - 3 y^2 \Big)  \Big(x_9^2 +y_9^2  -2\Gamma_9 \Big) - y_9  \big( xx_9 + yy_9 \big)  \frac{2 \Gamma_9 - x_9^2 - y_9^2 }{ \sqrt{\Gamma\big( 4\Gamma_9 - x_9^2 - y_9^2 \big)}}\Bigg)\\
\frac{\partial y_9}{\partial t} &= -\frac{3\,a^2\,\G\,m\,m_9}{16\,a_9^3\,\Gamma_9^2\,\epsilon_9^3} 
\Bigg( -2x_9 \bigg( xx_9 + yy_9 \bigg) \sqrt{\frac{4\Gamma_9 -x_9^2 - y_9^2}{\Gamma}} + x\bigg( 2\Gamma_9 - x_9^2 -y_9^2 \bigg) \sqrt{\frac{4\Gamma_9 - x_9^2 - y_9^2}{\Gamma}}  \\
&+\frac{1}{\Gamma}x_9\Big( 2 \Gamma - 3 x^2 -3y^2 \Big) \Big( x_9^2 + y_9^2 - 2\Gamma_9 \Big) - x_9 \Big( xx_9 + yy_9 \Big)  \frac{2\Gamma_9 - x_9^2 - y_9^2}{\sqrt{\Gamma\big(4\Gamma_9 -x_9^2 -y_9^2 \big)}} \Bigg) 
\end{align*}

The evolution of the sun's axial tilt is computed in the same manner. The Hamiltonian describing the cumulative effect of the planetary torques exerted onto the solar spin-axis is given by equation (\ref{HammySun}). Defining scaled Cartesian coordinates
\begin{align*}
\tilde{x}&=\sqrt{2\,\tilde{Z}}\cos{(\tilde{z})} & \tilde{y} &=\sqrt{2\,\tilde{Z}}\sin{(\tilde{z})},
\end{align*}  
we have:
\begin{align*}
\tilde{\mathcal{H}} &=\sum_{j}\left(\frac{\G\,m_j}{4a_j^3}\right)\tilde{a}^2 \bigg[\frac{3}{\tilde{\Gamma}}\bigg( \frac{\tilde{x}^2+\tilde{y}^2}{2} \bigg)  + \frac{3}{4} \sqrt{\frac{1}{\tilde{\Gamma}\Gamma}} (\tilde{x}x + \tilde{y}y) \bigg].
\end{align*}
Accordingly, Hamilton's equations are evaluated to characterize the dynamics of the sun's spin pole, under the influence of the planets:
\begin{align*}
\frac{d\tilde{x}}{dt} &= -\sum_{j} \left(\frac{\G m_j}{4a_j^3}\right)\tilde{a}^2 \Bigg( \frac{3}{4}y\sqrt{\frac{1}{\Gamma\tilde{\Gamma}}}+\frac{3\tilde{y}}{\tilde{\Gamma}}  \Bigg)\\
\frac{d\tilde{y}}{dt} &= \sum_{j}\left(\frac{\G m_j}{4a_j^3}\right) \tilde{a}^2 \Bigg( \frac{3}{4}x\sqrt{\frac{1}{\Gamma \tilde{\Gamma}}} +\frac{3 \tilde{x}}{\tilde{\Gamma}} \Bigg)
\end{align*}
Note that unlike $\Gamma$ and $\Gamma_9$, which are conserved, $\tilde{\Gamma}$ is an explicit function of time, and evolves according to the Skumanich relation. The above set of equations fully specifies the long-term evolution of the dynamical system.
\end{appendix}


\end{document}